\PassOptionsToPackage{dvipsnames,table}{xcolor}
\documentclass[conference,compsoc]{IEEEtran}
\IEEEoverridecommandlockouts
\usepackage{hyperref}
\usepackage{booktabs}
\usepackage{cite}
\usepackage{amsmath,amssymb,amsfonts}
\usepackage{algorithmic}
\usepackage{graphicx}
\usepackage{url}
\usepackage[export]{adjustbox} 
\usepackage{textcomp}
\usepackage{soul} 
\usepackage{url}
\usepackage[utf8]{inputenc}
\usepackage[T1]{fontenc}
\usepackage{fvextra}
\usepackage{xurl}        
\usepackage{hyperref}

\def\BibTeX{{\rm B\kern-.05em{\sc i\kern-.025em b}\kern-.08em
    T\kern-.1667em\lower.7ex\hbox{E}\kern-.125emX}}

\usepackage[many, minted]{tcolorbox}
\tcbuselibrary{minted}

\newtcblisting{mycode}{
  colback=gray!10,
  colframe=black,
  listing only,
  listing engine=minted,
  minted language=text,
  minted options={
    fontsize=\small,
    breaklines,
    breakanywhere,
    breaksymbolleft={},
    breaksymbolright={},
    breakautoindent=false,
    breakindent=0pt
  }
}

\begin{document}



\author{\IEEEauthorblockN{1\textsuperscript{st} Md Tanvirul Alam\IEEEauthorrefmark{1}\thanks{\IEEEauthorrefmark{1} Equal contribution. Work done while interning at Athena Security Group.}}
\IEEEauthorblockA{\textit{Rochester Institute of Technology} \\
Rochester, NY, USA \\
ma8235@rit.edu}
\and
\IEEEauthorblockN{2\textsuperscript{nd} Dipkamal Bhusal\IEEEauthorrefmark{1}}
\IEEEauthorblockA{\textit{Rochester Institute of Technology}\\
Rochester, NY, USA \\
db1702@rit.edu}
\and
\IEEEauthorblockN{3\textsuperscript{rd} Salman Ahmad}
\IEEEauthorblockA{\textit{Athena Security Group}\\
Jupiter, FL, USA \\
salman@athenasecuritygrp.com}
\and
\IEEEauthorblockN{4\textsuperscript{th} Nidhi Rastogi}
\IEEEauthorblockA{\textit{Rochester Institute of Technology}\\
Rochester, NY, USA \\
nxrvse@rit.edu}
\and
\IEEEauthorblockN{4\textsuperscript{th} Peter Worth}
\IEEEauthorblockA{\textit{Athena Security Group}\\
Jupiter, FL, USA \\
pworthjr@athenasecuritygrp.com}}

\title{AthenaBench: A Dynamic Benchmark for Evaluating LLMs in Cyber Threat Intelligence}

\maketitle
\thispagestyle{plain}
\pagestyle{plain}

\begin{abstract}
Large Language Models (LLMs) have demonstrated strong capabilities in natural language reasoning, yet their application to Cyber Threat Intelligence (CTI) remains limited. CTI analysis involves distilling large volumes of unstructured reports into actionable knowledge, a process where LLMs could substantially reduce analyst workload. CTIBench~\cite{alam2024ctibench} introduced a comprehensive benchmark for evaluating LLMs across multiple CTI tasks. In this work, we extend CTIBench by developing \textit{AthenaBench}, an enhanced benchmark that includes an improved dataset creation pipeline, duplicate removal, refined evaluation metrics, and a new task focused on risk mitigation strategies. We evaluate twelve LLMs, including state-of-the-art proprietary models such as GPT-5 and Gemini-2.5 Pro, alongside seven open-source models from the LLaMA and Qwen families. While proprietary LLMs achieve stronger results overall, their performance remains subpar on reasoning-intensive tasks, such as threat actor attribution and risk mitigation, with open-source models trailing even further behind. These findings highlight fundamental limitations in the reasoning capabilities of current LLMs and underscore the need for models explicitly tailored to CTI workflows and automation.
\end{abstract}


\section{Introduction}

Large Language Models (LLMs) such as ChatGPT~\cite{ouyang2022training}, LLaMA~\cite{touvron2023llama}, and their specialized variants have demonstrated remarkable proficiency in language understanding, reasoning, and knowledge synthesis~\cite{bubeck2023sparks}. Their ability to process and integrate vast textual corpora has driven advances across domains such as education, healthcare, and scientific discovery. Among emerging applications, Cyber Threat Intelligence (CTI) presents a particularly promising yet underexplored area. CTI involves collecting, analyzing, and disseminating information about current and emerging cyber threats to help organizations anticipate and mitigate attacks~\cite{mcmillan2013definition}. Effective CTI analysis demands processing large volumes of unstructured text—such as vulnerability disclosures, malware reports, and threat campaign summaries—and transforming them into actionable insights. This process requires factual recall, contextual comprehension, and reasoning to infer hidden relationships among vulnerabilities, threat actors, and attack techniques~\cite{alaeifar2024current}.

Early studies suggest that LLMs could assist or accelerate many CTI workflows, ranging from drafting threat summaries to mapping observed indicators to known adversaries or techniques~\cite{zaboli2024chatgpt, yigit2024review, fieblinger2024actionable}. Both open-source and proprietary models have shown promise on CTI-related tasks through prompt engineering and few-shot demonstrations. However, these initial results fail to capture the breadth required to evaluate LLM capabilities in a domain as complex and nuanced as CTI, which encompasses a wide variety of interrelated aspects. Without standardized benchmarks, it remains challenging to assess model performance or rigorously compare progress across systems comprehensively.

Existing evaluation efforts often fail to capture the practical aspects of cybersecurity, as they design tasks that primarily test the memorization ability of LLMs rather than measuring applied security reasoning~\cite{li2023seceval, liu2024cyberbench, liu2023secqa, ji2024sevenllm}. Knowledge-based benchmarks such as SECURE~\cite{bhusal2024secure} and CTIBench~\cite{alam2024ctibench} attempt to address these limitations by introducing knowledge-intensive tasks that emphasize security reasoning. However, they suffer from static knowledge issues, since the benchmark tasks are derived from fixed corpora. Given that cybersecurity is a rapidly evolving domain, benchmarks must be continuously updated to measure LLM reasoning performance~\cite{mallick2024navigating} reliably. Without such updates, benchmarks risk becoming outdated and misaligned with the evolving cyber threat landscape.

\begin{figure}
    \centering
    \includegraphics[width=0.98\linewidth]{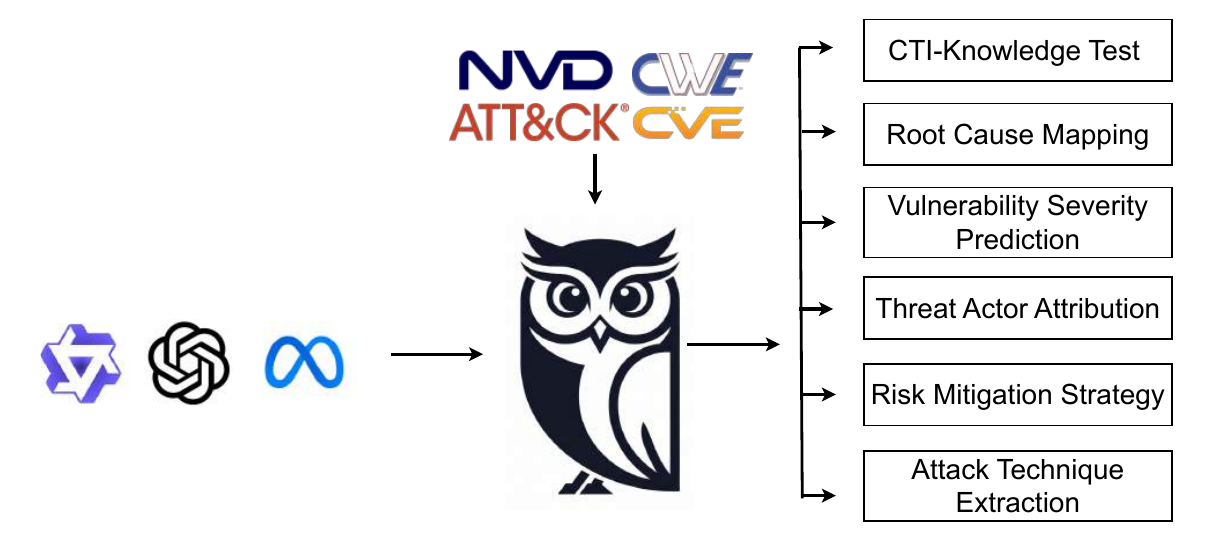}
    \caption{Overview of LLM benchmarking tasks in AthenaBench}
    \label{fig:placeholder}
\end{figure}

To address these limitations, we introduce \texttt{AthenaBench}, a dynamic benchmark suite designed to evaluate LLMs on realistic, knowledge-intensive CTI tasks. AthenaBench builds upon prior efforts~\cite{alam2024ctibench} in three key ways:

\begin{itemize}
    \item \textbf{Improved Task and Dataset Design.} We expand CTIBench~\cite{alam2024ctibench} by introducing a novel task, \emph{Risk Mitigation Strategies (RMS)}, which evaluates a model’s ability to propose effective defensive measures. In addition, we refine existing tasks by removing duplicates in vulnerability-related datasets, reducing contamination risks, and enhancing tasks such as threat actor attribution and attack technique extraction. We also collect a larger and more up-to-date set of examples for each task to better reflect current CTI practices.
    
    \item \textbf{Dynamic Data Construction.} Unlike benchmarks that rely on static corpora, AthenaBench leverages live CTI data sources and APIs, including MITRE ATT\&CK~\cite{strom2018mitre} and the NVD API~\cite{NVDDatabase}, to continuously generate benchmark samples. This enables tasks to remain current with minimal human supervision, ensuring that models are evaluated on emerging vulnerabilities, techniques, and threat actors in alignment with real-world developments.
    
    \item \textbf{Enhanced Evaluation Metrics.} We improve task-specific evaluation metrics and introduce a unified aggregated score that captures performance across all tasks. This provides a comprehensive lens for comparing the CTI reasoning capabilities of the models and highlights their relative strengths and weaknesses.
\end{itemize}

We evaluate twelve proprietary and open-source LLMs: GPT-4~\cite{gpt4}, GPT-4o~\cite{gpt4o}, GPT-5~\cite{gpt5}, Gemini-2.5 (Flash and Pro)~\cite{gemini2.5}, Qwen3 (4B, 8B, 14B)~\cite{qwen3technicalreport}, Llama 3.1-8B~\cite{llama38b}, Llama 3-70B~\cite{llama370b}, Llama 3.3-70B~\cite{llama3.370b} and Llama-Primus-Merged~\cite{llamaprimus} across all AthenaBench tasks. Our results show that proprietary models achieve moderate success, with GPT-5 attaining the highest overall score, followed by Gemini-2.5 Pro, while open-source models lag significantly behind. Performance also varies notably across tasks: models perform well on structured prediction tasks like Vulnerability Severity Prediction (VSP) but struggle on reasoning-intensive ones, highlighting persistent challenges in achieving full automation of CTI workflows. The code and benchmark dataset are publicly available at \url{https://github.com/Athena-Software-Group/athenabench}.

\section{Background and related work}

Large Language Models (LLMs) have rapidly transformed the landscape of natural language processing, enabling advances in reasoning, information synthesis, and code generation~\cite{bubeck2023sparks}. Recently, several open models have been adapted for cybersecurity use cases, including vulnerability detection~\cite{shestov2024finetuning, ferrag2023securefalcon, fang2024llm, li2024llm}, IT operations automation~\cite{guo2023owl}, security knowledge assistance~\cite{sultana2023towards}, and knowledge graph construction~\cite{fieblinger2024actionable, huang2024ctikg}. These efforts underscore the increasing potential of LLMs to enhance security analysis workflows. However, existing evaluation efforts primarily focus on narrow downstream tasks and do not adequately assess the higher-level reasoning capabilities required for real-world security investigations.

While general-purpose benchmarks such as GLUE~\cite{wang2018glue}, MMLU~\cite{hendrycks2020measuring}, and KOLA~\cite{yu2023kola} broadly evaluate language model capabilities, they do not specifically target reasoning in cybersecurity domains. Cybersecurity-specific benchmarks such as CyberMetric~\cite{tihanyi2024cybermetric}, CyberBench~\cite{liu2024cyberbench}, SecEval~\cite{li2023seceval}, and SecQA~\cite{liu2023secqa} primarily measure memorization, often using multiple-choice questions about cybersecurity concepts. Other benchmarks, including NetEval~\cite{miao2023empirical} and OpsEval~\cite{liu2023opseval}, focus on operational IT tasks such as alert triage and root cause analysis. In the Cyber Threat Intelligence (CTI) domain specifically, CTIBench \cite{alam2024ctibench} and SECURE \cite{bhusal2024secure} evaluate LLMs on knowledge-intensive CTI tasks. However, these benchmarks rely on static datasets that quickly become outdated and fail to reflect the fast-evolving threat landscape. AthenaBench addresses these gaps by introducing a suite of knowledge-intensive CTI tasks that are continuously updated through live integration with authoritative CTI frameworks and threat reporting sources.

\section{Methodology: Benchmark Design}
To evaluate the effectiveness of large language models (LLMs) in understanding cyber threat intelligence (CTI), we design six complementary evaluation tasks. These tasks target distinct dimensions of CTI reasoning: \emph{recalling factual knowledge}, \emph{predicting vulnerability severity}, \emph{mapping root causes of vulnerabilities}, \emph{attributing threats}, \emph{prescribing defenses}, and \emph{extracting attack techniques}. The datasets are dynamically populated from up-to-date threat intelligence sources and live security frameworks, including MITRE ATT\&CK~\cite{strom2018mitre}, the NVD API~\cite{NVDDatabase}, and recent advanced persistent threat (APT) reports. This design enables evaluation not only of whether a model has memorized legacy information, but also of its ability to interpret, analyze, and reason about \emph{emerging} threats in real-world contexts.

\subsection{CTI Knowledge Test (CKT)}
\textbf{Objective.} The CTI Knowledge Test (CKT) evaluates an LLM’s ability to demonstrate foundational knowledge in cyber threat intelligence (CTI) across key frameworks, concepts, and best practices. This includes familiarity with the behaviors, attack techniques, defensive strategies, regulatory frameworks, and standards for threat intelligence. Such breadth is essential for simulating the reasoning challenges faced by security analysts in day-to-day operations.

\textbf{Data construction.} We curate content from authoritative Cyber Threat Intelligence (CTI) sources, including the MITRE ATT\&CK framework~\cite{strom2018mitre}, the CWE~\cite{christey2013common} and CAPEC~\cite{capec} repositories, CISA advisories~\cite{cisa_advisories}, standards such as STIX/TAXII~\cite{OASIS}, and regulatory frameworks such as GDPR~\cite{GDPR}. Following prior work~\cite{alam2024ctibench}, we utilize GPT-5 to generate multiple-choice questions (MCQs) directly from this curated corpus. The prompting strategy ensures that each question contains exactly five answer options with only one correct choice and is designed to match the knowledge level of a CTI professional with approximately 3–5 years of experience. The number of questions generated per source scales proportionally with the length of the original document, resulting in a total of 3,879 questions.


\textbf{Validation.} We employed a two-stage validation pipeline to ensure dataset quality. In the first stage, two state-of-the-art models, GPT-5 and Gemini Pro, independently attempted each question, and instances where either model failed were flagged for review. In the second stage, human evaluators manually inspected all flagged cases, removing 279 ambiguous or contextually insufficient questions. In addition, 126 questions with corrected answer options were retained to fix response errors and improve clarity and formatting.

\textbf{Dataset.} After validation, we sampled 3,000 questions to construct the final CTI Knowledge Test dataset comprising high-quality multiple-choice questions spanning multiple CTI knowledge domains.

\textbf{Task.} Each CKT benchmark instance consists of an MCQ with five options and a ground-truth correct answer.

\begin{tcolorbox}[title= CKT Example]
\textbf{Question:} What is the most appropriate containment-focused action if telemetry shows KONNI modifying ComSysApp to load a malicious DLL and registering itself as a Windows service?

\textbf{Options:} A) Immediately rotate all browser-stored credentials enterprise-wide, B) Baseline clipboard access to detect future data theft, C) Disable and audit ComSysApp and related service registrations; block malicious DLL via AppLocker/SRPs, D) Isolate only endpoints showing net session usage, E) Focus on decrypting KONNI C2 by disabling TLS interception.

\vspace{1em}
\textbf{Answer:} C
\end{tcolorbox}

\subsection{Root Cause Mapping (RCM)}
\textbf{Objective.}  
The Root Cause Mapping (RCM) task evaluates an LLM’s ability to identify the underlying cause of a vulnerability by mapping its natural language description to the correct Common Weakness Enumeration (CWE) category~\cite{christey2013common}. This capability is critical for vulnerability triage and for linking newly reported CVEs to established classes of weaknesses.

\textbf{Data construction.}  
We construct the RCM dataset using records from the National Vulnerability Database (NVD) via its JSON API (v2.0)~\cite{NVDDatabase}. All CVE entries within a configurable publication window are retrieved through authenticated API requests. For each CVE, we extract its English description, publication date, last modified timestamp, and any linked CWE identifiers from the \texttt{weaknesses} field. To ensure label clarity, we retain only vulnerabilities annotated with exactly one CWE. To improve dataset quality, we deduplicate entries with identical descriptions (keeping the latest record when duplicates exist) and remove descriptions that are too short (less than 25 words) to provide sufficient semantic content.

\textbf{Dataset.} Using a publication window from \texttt{2025-01-01} to \texttt{2025-07-31}, we construct a dataset of 2,000 high-quality RCM instances.

\textbf{Task.}  
Each benchmark instance consists of a vulnerability description as the prompt and the correct CWE identifier as the ground-truth label.

\begin{tcolorbox}[title= RCM Example]
\textbf{CVE Description:} The Ultimate Store Kit Elementor Addons, Woocommerce Builder, EDD Builder, Elementor Store Builder, Product Grid, Product Table, Woocommerce Slider plugin for WordPress is vulnerable to Cross-Site Request Forgery in all versions up to, and including, 2.4.1. This is due to missing or incorrect nonce validation on the dismiss() function. This makes it possible for unauthenticated attackers to set arbitrary user meta values to `1` which can be leveraged to lock and administrator out of their site via a forged request granted they can trick a site administrator into performing an action such as clicking on a link.

\vspace{1em}
\textbf{Answer:} CWE-352
\end{tcolorbox}

\subsection{Vulnerability Severity Prediction (VSP)}
\textbf{Objective.}  
The Vulnerability Severity Prediction (VSP) task evaluates an LLM’s ability to assess the severity of a vulnerability from its natural language description by predicting its CVSS v3.1 base vector. This probes whether a model can infer technical impact (e.g., on confidentiality, integrity, and availability consequences).

\textbf{Data construction.}  
For each CVE, we parse the \texttt{metrics} field to extract CVSS v3.1 and v4.0 data, distinguishing between NVD-assigned and CNA-assigned metrics. When both are available, NVD metrics are preferred. Each record includes the CVSS vector, base score, and severity level, as well as its natural language description and CWE tags.  

To ensure dataset consistency, we deduplicate descriptions, filter out entries that are very short (less than 30 words), and retain only those vulnerabilities with valid CVSS v3.1 vectors. 

\textbf{Dataset.}  
Using the publication window from \texttt{2025-01-01} to \texttt{2025-07-31}, we construct a dataset of 2,000 high-quality VSP instances.

\textbf{Task.}  
We frame VSP as a text-to-structured prediction problem. Each benchmark instance consists of a vulnerability description as the prompt and its corresponding CVSS v3.1 vector as the target.  

\begin{tcolorbox}[title= VSP Example]
\textbf{CVE Description:} A command injection vulnerability exists in the workflow-checker.yml workflow of significant-gravitas/autogpt. The untrusted user input `github.head.ref` is used insecurely, allowing an attacker to inject arbitrary commands. This vulnerability affects versions up to and including the latest version. An attacker can exploit this by creating a branch name with a malicious payload and opening a pull request, potentially leading to reverse shell access or theft of sensitive tokens and keys.

\vspace{1em}
\textbf{Answer:} 

CVSS:3.1/AV:N/AC:L/PR:N/UI:N/S:U/C:H/I:H/A:H
\end{tcolorbox}

\subsection{Threat Actor Attribution (TAA)}
\textbf{Objective.}  
The Threat Actor Attribution (TAA) task evaluates a model’s ability to perform abductive attribution, inferring which adversary is responsible for an observed pattern of activity. This mirrors the core analytical challenge faced by CTI analysts, who must connect fragmentary behavioral evidence to known threat groups without relying on explicit name mentions.

\textbf{Data construction.}  
We curate a corpus of public threat intelligence reports from vendors and security blogs that describe real-world APT operations, such as Malpedia~\cite{malopedia}. Each report is automatically parsed using a hybrid pipeline combining \texttt{newspaper3k}\footnote{\url{https://newspaper.readthedocs.io/en/latest/}} and BeautifulSoup\footnote{\url{https://pypi.org/project/beautifulsoup4/}}. GPT-5 is then prompted with a carefully designed instruction 
(see Appendix~\ref{appendix:taainstruction}) 
that rewrites the raw prose into a concise, bullet-pointed profile of behaviors, replacing actor names with pronouns such as \emph{they}.


\textbf{Human validation.}  
The anonymization process occasionally leaves subtle identifiers that leak the answer. For example, a report about APT28 could be anonymized but still mention “the Russian GRU hacking group,” which would reveal the actor. To mitigate this risk, all generated instances are manually reviewed by human annotators.

\textbf{Dataset.} After validation, the final dataset comprises 100 high-quality instances of threat actor attribution.

\textbf{Task.}  
Each benchmark instance includes the anonymized description as input, and the correct actor name as the target answer.

\begin{tcolorbox}[title= TAA Example]
\textbf{Threat report:} 
- They use creative malware campaigns leveraging Google Workspace apps, including Google Sheets and Google Drive, for command-and-control. - They used malware families including VOLDEMORT, DUSTTRAP, and TOUGHPROGRESS. - They have used public cloud and free web hosting for malware distribution and C2. - Since at least August 2024, they have used free web hosting tools to distribute malware, including VOLDEMORT, DUSTTRAP, and TOUGHPROGRESS. - They frequently used Cloudflare Workers subdomains for hosting malware. - They also used InfinityFree and TryCloudflare hosting services. - They distributed phishing links via URL shorteners that redirect to malware on free hosting subdomains. - Example Cloudflare Workers subdomains used: word.msapp.workers.dev, cloud.msapp.workers.dev. - Example TryCloudflare domains used: term-restore-satisfied-hence.trycloudflare.com, ways-sms-pmc-shareholders.trycloudflare.com. - Example InfinityFree domains used: resource.infinityfreeapp.com, pubs.infinityfreeapp.com. - Example URL shorteners used: lihi.cc, tinyurl.com, my5353.com, reurl.cc. - Targets included hundreds of recipients across various geographies and industries. - Activity observed and discussed includes campaigns active through at least October 2024. - Indicators of compromise related to these campaigns are available via Google Threat Intelligence."

\vspace{1em}
\textbf{Answer:} APT41

\end{tcolorbox}

\subsection{Risk Mitigation Strategy (RMS)}
\textbf{Objective.} The RMS task evaluates whether LLMs can recommend appropriate MITRE ATT\&CK mitigation strategies \cite{mitremitigations} based on a detailed description of an observed attack scenario. This tests the model’s ability to apply structured cybersecurity knowledge to real-world cases and to reason about defensive countermeasures.

\textbf{Data Collection.} We construct this dataset using the official MITRE ATT\&CK STIX 2.0 enterprise bundle, retrieved from MITRE’s public repository\footnote{\url{https://raw.githubusercontent.com/mitre/cti/master/enterprise-attack/enterprise-attack.json}}. Each attack-pattern object is first linked to its corresponding course-of-action (mitigation) objects through mitigates relationship edges. To promote recency and prevent training-set leakage, we apply a configurable time window filter using the created or modified timestamps embedded in the STIX objects, ensuring only techniques updated within the specified start–end range are included.

To make the task realistic, we convert each technique’s textual description into a brief, specific, 2–3 sentence attack scenario. These scenarios are automatically generated using GPT-5 with carefully designed instructions and few-shot exemplars that demonstrate the desired style (see Appendix~\ref{appendix:systeminstructrms}). This forces the model to recognize the underlying technique from behavioral context rather than keyword matching. Each sample thus consists of a tuple: (attack scenario, correct mitigation set).

\textbf{Dataset.} With start and end date of \texttt{2024-01-01} and \texttt{2025-07-31}, we create 500 questions for RMS dataset.

\textbf{Task.} For each selected technique, we produce a benchmark entry where the input is the GPT-generated scenario and the expected answer is the set of linked mitigation IDs (M10xx) from MITRE ATT\&CK.

\begin{tcolorbox}[title= RMS Example]
\textbf{Attack description:} After obtaining local admin credentials on a Windows host with Task Scheduler running, the adversary used the at command to schedule cmd.exe to launch a dropped payload under the SYSTEM account at 03:00. They then repeated the technique via the Win32\_ScheduledJob WMI class to create a second job later that day to ensure recurring execution.

\vspace{1em}
\textbf{Answer:} M1018, M1026, M1028, M1047
\end{tcolorbox}

\subsection{Attack Technique Extraction (ATE)}
\textbf{Objective.} ATE tests whether LLMs can identify the correct MITRE ATT\&CK technique ID \cite{mitreattack} underlying a realistic attack scenario. While RMS focuses on mitigation recommendation, ATE targets the upstream cognitive skill of recognizing the offensive technique itself from its behavioral description.

\textbf{Data Collection.} We reuse the same data pipeline as RMS to ensure consistency. Attack patterns are obtained from the MITRE ATT\&CK STIX bundle within the specified time window and filtered to include only active, non-deprecated techniques. We again use GPT-5 to transform each raw technique description into a realistic multi-sentence scenario. This indirect formulation prevents models from exploiting memorized surface phrases from MITRE and requires reasoning about the semantics of the behavior.

\textbf{Dataset.} With start and end date of \texttt{2024-01-01} and \texttt{2025-07-31}, we create 500 questions for ATE dataset. 

\textbf{Task.} Each benchmark entry contains the scenario text and an environment hint, and the correct Txxxx ID.

\begin{tcolorbox}[title= ATE Example]
\textbf{Attack description:} After gaining admin access to a Windows server, the attacker stopped the endpoint protection service and modified its configuration file to prevent real-time scanning from starting on boot. They then added their working directory to the EDR exclusion list and changed the Sysmon Autologger “Start” and “Enable” registry values to suppress event collection before deploying their payload.

\vspace{1em}
\textbf{Answer:} T1562
\end{tcolorbox}

\section{Experiments and Results}

\subsection{Experimental Settings}
We evaluate twelve proprietary and open-source LLMs: GPT-4~\cite{gpt4}, GPT-4o~\cite{gpt4o}, GPT-5~\cite{gpt5}, Gemini-2.5 (Flash and Pro)~\cite{gemini2.5}, Qwen3 (4B, 8B, 14B)~\cite{qwen3technicalreport}, Llama 3.1-8B~\cite{llama38b}, Llama 3-70B~\cite{llama370b}, Llama 3.3-70B~\cite{llama3.370b} and Llama-Primus-Merged~\cite{llamaprimus} across all AthenaBench tasks. Llama-Primus-Merged is finetuned on cybersecurity datasets, while others are general-purpose LLMs. We set the decoding temperature to 0 for all models except GPT-5, which does not expose a temperature parameter. For GPT-5, we use its default reasoning configuration. Appendix~\ref{appendix:evalprompts} provides the evaluation prompts used across all tasks.

\begin{table*}[]
\centering
\caption{Performance of proprietary and open-source LLMs across AthenaBench tasks. Each value represents the model’s performance on the full benchmark, while values in parentheses indicate results on the publicly released mini-subset. “Acc” denotes accuracy, and “F1” denotes F1-score. Higher values indicate better performance across all tasks.}
\label{tab:my-table}
\resizebox{0.8\textwidth}{!}{%
\begin{tabular}{@{}lccccccc@{}}
\toprule
\multicolumn{1}{c}{\textbf{Model}} &
  \textbf{CKT (Acc)} &
  \textbf{ATE (Acc)} &
  \textbf{RCM (Acc)} &
  \textbf{RMS (F1)} &
  \textbf{VSP (Acc)} &
  \textbf{TAA (Acc)} &
  \textbf{Combined} \\ \midrule
GPT-4                  & 78.7 (80.3) & 35.8 (43.0) & 63.1 (60.0) & 15.1 (13.7) & 84.7 (84.2) & 31.0 (26.0) & 51.4 (51.2) \\
GPT-4o                 & 85.2 (87.7) & 51.6 (59.0) & 71.3 (68.0) & 20.2 (19.9) & 84.7 (85.9) & 35.0 (\textbf{30.0}) & 58.0 (58.4) \\
GPT-5                  & \textbf{92.0} (\textbf{96.0}) & 76.0 (\textbf{77.0}) & \textbf{71.6} (\textbf{69.5}) & \textbf{32.6} (\textbf{33.0}) & \textbf{85.4} (\textbf{88.3}) & \textbf{39.0} (\textbf{30.0}) & \textbf{66.1} (\textbf{65.6}) \\
Gemini-2.5-flash       & 85.1 (87.7) & 51.6 (57.0) & 65.1 (64.5) & 13.4 (14.0) & 78.5 (78.3) & 30.0 (22.0) & 54.0 (53.9) \\
Gemini-2.5-pro         & 89.1 (91.0) & \textbf{76.2} (\textbf{77.0}) & 71.2 (68.0) & 28.4 (29.0) & \textbf{85.4} (86.7) & 31.0 (24.0) & 63.6 (62.6) \\ \midrule
Qwen3-4B               & 74.7 (76.3) & 5.6 (8.0) & 45.4 (43.5) & 4.8 (5.8) & 79.6 (78.2) & 15.0 (16.0) & 37.5 (38.0) \\
Qwen3-8B               & 75.7 (75.3) & 11.8 (13.0) & 48.9 (45.5) & 5.5 (6.8) & 82.6 (82.6) & 16.0 (20.0) & 40.1 (40.5) \\
Qwen3-14B              & 78.6 (82.7) & 19.4 (21.0) & 54.1 (49.0) & 7.0 (8.5) & 80.3 (78.0) & 17.0 (16.0) & 42.7 (42.5) \\
Llama 3.1-8B           & 71.8 (74.0) & 16.4 (16.0) & 42.8 (41.0) & 3.6 (5.4) & 74.0 (74.1) & 24.0 (24.0) & 38.8 (39.1) \\
Llama 3-70b-Instruct   & 78.9 (81.0) & 31.6 (37.0) & 56.7 (54.5) & 11.1 (10.9) & 63.8 (63.4) & 22.0 (24.0) & 44.0 (45.1) \\
Llama 3.3-70b-Instruct & 81.4 (81.7) & 30.4 (44.0) & 60.0 (59.0) & 11.1 (11.5) & 70.1 (69.7) & 26.0 (22.0) & 46.5 (48.0) \\
Llama-Primus-Merged    & 76.3 (79.7) & 33.8 (32.0) & 56.6 (51.0) & 6.6 (6.4) & 71.9 (71.8) & 17.0 (18.0) & 43.7 (43.1) \\ \bottomrule
\end{tabular}%
}
\end{table*}

\subsection{Evaluation metrics}  
For \emph{CTI Knowledge Test (CKT)}, \emph{Root Cause Mapping (RCM)}, and \emph{Threat Actor Attribution (TAA)}, we use accuracy as the primary evaluation metric, since each task has a single correct categorical answer. For \emph{Risk Mitigation Strategy (RMS)}, the evaluation involves multi-label classification, where each attack scenario may correspond to multiple valid mitigation strategies; therefore, we employ the F1-score metric.  

For \emph{Vulnerability Severity Prediction (VSP)}, although models are prompted to predict the CVSS vector string, we derive the corresponding numerical CVSS score using the Python \texttt{cvss} library~\cite{cvsspackage}. We then compute the Mean Absolute Deviation (MAD) between model-predicted and ground-truth severity scores. Because the observed scores in our dataset range from 2.3 to 10, we normalize MAD over this interval and convert it into a percentage-based accuracy metric for consistency with other tasks:

\begin{equation*}
\mathrm{Accuracy}_{\mathrm{VSP}} = 1 - \frac{\mathrm{MAD}}{7.7}
\end{equation*}

This formulation ensures that all AthenaBench tasks are evaluated on a uniform percentage scale, facilitating the meaningful aggregation of results across tasks.

\subsection{AthenaBench-Mini}
To support ongoing research and enable faster model evaluation, we release a compact version of the benchmark, \texttt{AthenaBench-Mini}, which is publicly available\footnote{\url{https://github.com/Athena-Software-Group/athenabench}}. This subset contains 10\% of the original samples for the CKT, RCM, and VSP tasks, 20\% for the RMS and ATE tasks, and 50\% for the TAA tasks. We generated multiple candidate subsets using ten random seeds and selected the one whose aggregate performance most closely aligned with the complete benchmark. Upon paper acceptance, we will introduce a formal process for researchers to request non-commercial access to the complete AthenaBench dataset.

Table~\ref{tab:athenabench-sizes} provides statistics on the number of samples across different tasks in AthenaBench for both the Full and Mini subsets.

\begin{table}[h!]
\centering
\caption{Number of samples across AthenaBench tasks for the Full and Mini subsets.}
\label{tab:athenabench-sizes}
\resizebox{0.85\linewidth}{!}{
\begin{tabular}{lcccccc}
\toprule
\textbf{Dataset} & \textbf{CKT} & \textbf{ATE} & \textbf{RCM} & \textbf{RMS} & \textbf{VSP} & \textbf{TAA} \\ 
\midrule
Full & 3000 & 500 & 2000 & 500 & 2000 & 100 \\
Mini & 300 & 100 & 200 & 100 & 200 & 50 \\
\bottomrule
\end{tabular}
}
\end{table}

\subsection{Results}
Table~\ref{tab:my-table} summarizes the performance of proprietary and open-source LLMs across all six AthenaBench tasks. Each cell reports results on the full benchmark, with values in parentheses indicating performance on the mini subset. While minor variations in per-task rankings are observed between the full and mini versions of the benchmark, the overall model rankings remain consistent when evaluated using the combined aggregate metric.

\textbf{CTI Knowledge Test (CKT).} CKT represents the most basic factual knowledge retrieval task, focused on cybersecurity frameworks, standards, and definitions. As expected, all large proprietary models exceed 80\% accuracy, with GPT-5 achieving 92.0\%. These results indicate that advanced models exhibit strong factual recall across CTI concepts, whereas open-source models perform notably worse.

\textbf{Attack Technique Extraction (ATE).} This task, which involves mapping realistic attack scenarios to MITRE ATT\&CK technique IDs, remains particularly challenging. GPT-5 and Gemini-2.5 Pro achieve accuracies around 76\%, while GPT-4 and GPT-4o lag, and open-source models perform substantially worse. These results highlight that reasoning over behavioral patterns remains a challenging capability for current models.

\textbf{Root Cause Mapping (RCM).} In this vulnerability-to-CWE mapping task, proprietary models cluster around 71\% accuracy, demonstrating consistent ability to associate natural language vulnerability descriptions with corresponding weakness categories. Open-source models, however, show markedly lower performance.

\textbf{Risk Mitigation Strategy (RMS).} RMS emerges as one of the most challenging tasks. Even top models, GPT-5 (32.6\%) and Gemini-2.5 Pro (28.4\%), achieve only modest F1-scores, with GPT-4 and GPT-4o performing worse. Open-source models fare significantly lower. The task requires reasoning about defensive measures in nuanced scenarios, underscoring current limits of applied CTI reasoning.

\textbf{Vulnerability Severity Prediction (VSP).} Unlike RMS, VSP exhibits uniformly high accuracy. Since CVSS scoring is structured and standardized, models can reliably map descriptions to severity vectors once trained on enough examples. 


\textbf{Threat Actor Attribution (TAA).} All models achieve low accuracy on TAA, with GPT-5 performing best at 39\% and GPT-4o following at 35\%. Open-source models remain notably weaker. Attribution demands abductive reasoning to infer the responsible actor from limited behavioral evidence, underscoring a core challenge for current LLMs.

\textbf{Overall performance.} Proprietary reasoning-optimized models (GPT-5, Gemini-2.5-pro, GPT-4o) consistently outperform both earlier-generation proprietary models (GPT-4) and open-source models. GPT-5 achieves the highest combined score (66.1\%), demonstrating more balanced reasoning across tasks. Gemini-2.5-pro follows closely (63.6\%). In contrast, even large open-source models such as LLaMA-3.3-70B achieve only 46.5\% on average. Llama-Primus-Merged, which was post-trained on Llama 3.1-8B, outperforms the base model by 4.9 percentage points, underscoring the effectiveness of domain-specific instruction tuning for enhancing LLM reasoning in CTI-related tasks.

Analysis of LLMs on AthenaBench reveals that structured tasks (e.g., VSP, RCM) are tractable for many models, while open-ended applied reasoning tasks (e.g., RMS, TAA, ATE) expose significant weaknesses. Proprietary frontier models show meaningful advances, but their performance is still far from reliable automation of CTI workflows. Open-source models, despite rapid scaling, remain brittle outside narrow structured domains.

\subsection{Results with Web Search Capability}

The two vulnerability-focused tasks, RCM and VSP, in AthenaBench evaluate model performance on newly discovered vulnerabilities that lie beyond training data and rely on source descriptions such as NVD entries, making them unsuitable for web search evaluation. In contrast, the ATE, RMS, and TAA tasks can leverage web search capabilities. These tasks were designed by paraphrasing and reformulating the source descriptions to minimize the likelihood of direct answer retrieval through simple search. Table~\ref{tab:websearch_results} summarizes the results, with values in parentheses indicating percentage improvements over the non-search baseline.

\begin{table}[h!]
\centering
\caption{Performance of GPT-5 with Web Search Capability}
\label{tab:websearch_results}
\begin{tabular}{lccc}
\hline
\textbf{Model} & \textbf{ATE (Acc)} & \textbf{RMS (F1)} & \textbf{TAA (Acc)} \\ 
\hline
GPT-5-Search & 79.6 (+3.6) & 51.9 (19.3) & 63.0 (+24.0) \\ 
\hline
\end{tabular}
\end{table}

As shown in Table~\ref{tab:websearch_results}, the GPT-5 model exhibits consistent performance gains across all three tasks, with particularly notable improvements in the RMS and TAA tasks, highlighting the effectiveness of web search integration in enhancing factual retrieval and reasoning.

Below we show an example question from the RMS task where the base GPT-5 model produced an incorrect answer, whereas GPT-5 with web search was able to identify the correct mitigation response.

\begin{tcolorbox}[title=Attack Scenario]
After gaining brief physical access to a target workstation, the adversary connected a TinyPilot KVM device inline between the PC and its peripherals. They then used the device's IP-based interface to establish an interactive remote session, issuing keyboard and mouse inputs to control the system without triggering host-based security tools.
\end{tcolorbox}

\begin{tcolorbox}[title=GPT-5 response]
M1030 Network Segmentation limits or blocks network reachability to unauthorized devices and management interfaces. By isolating or denying connectivity to rogue inline KVMs (e.g., TinyPilot) on user segments, their IP-based control interfaces cannot be reached, preventing remote keyboard/mouse injection despite physical insertion.

Answer: M1030
\end{tcolorbox}

\begin{tcolorbox}[title=GPT-5 with search response]
M1034 Limit Hardware Installation best fits: ATT\&CK explicitly recommends blocking unapproved peripherals and IP-based KVM/remote access hardware to prevent abuses of hardware additions and inline KVM devices like TinyPilot. (\url{https://attack.mitre.org/techniques/T1219/003/?utm_source=openai})

Answer: M1034
\end{tcolorbox}

\section{Limitations}
AthenaBench provides a representative subset of tasks to evaluate LLMs’ reasoning capabilities in cyber threat intelligence (CTI). However, it does not fully capture the end-to-end complexity of real-world CTI operations. Expanding the task set to include additional analytic and operational activities remains an important direction for future work. The current datasets also rely primarily on English-language sources, excluding a substantial portion of intelligence published in other languages. Some tasks depend on GPT-5 for data generation, which may confer a modest advantage to that model~\cite{panickssery2024llm}. Finally, while AthenaBench emphasizes reasoning- and knowledge-intensive evaluation, it does not yet address operational aspects such as efficiency, latency, and robustness under adversarial conditions. These dimensions are crucial for practical CTI deployment and will be incorporated into future iterations of the benchmark.

\section{Conclusion}
In this work, we introduced \textit{AthenaBench}, a dynamic benchmark suite for evaluating LLMs on realistic, knowledge-intensive cyber threat intelligence (CTI) tasks. By integrating live sources such as the MITRE ATT\&CK framework and the NVD vulnerability database, AthenaBench ensures that benchmark samples evolve with the threat landscape, offering a more faithful measure of models’ applied reasoning in CTI. Our results show that while reasoning-optimized models like GPT-5 and Gemini-2.5 Pro demonstrate strong progress, notable gaps remain in reasoning-intensive tasks such as attribution and mitigation strategy generation. We hope AthenaBench serves as a foundation for systematic evaluation and continuous advancement of LLMs in cybersecurity research and automation.

\clearpage
\newpage

\bibliographystyle{IEEEtran}
\bibliography{IEEEabrv,ref}

\clearpage
\appendix

\subsection{CTI Knowledge Test System Instruction Promot}\label{appendix:ctipromptsystem}
\begin{tcolorbox}[title= CKT System Instruction]
You are a cybersecurity expert specializing in Cyber Threat Intelligence (CTI). Given the text below, generate a maximum of MAX multiple-choice questions with five possible options. Follow these strict rules:
1. Each question must have exactly 5 answer options (A–E).
2. Each question must be non-trivial and require 3–5 years of CTI experience to answer. Include a balanced mix of rote recall (definitions, key facts, sub-technique identification) and 
operational judgement.
3. Use correct technical terms, identifiers, and context from the source.
5. Each question must have only ONE correct answer.
6. Output must be in TSV format with the following columns: Question, Option A, Option B, Option C, Option D, Option E, Correct Answer(A/B/C/D/E), Explanation.

Do not include any commentary or text outside the TSV table.

Text: \{\}
\end{tcolorbox}

\subsection{Threat Actor Attribution System Instruction}\label{appendix:taainstruction}

\begin{tcolorbox}[title= TAA System Instruction]
You are a cybersecurity assistant. The following report text is about a specific threat actor. Extract only factual details about their activities as a bullet-point list. Rules: 1) Replace the primary threat actor's name with 'they'. 2) If any other threat actor is mentioned, OMIT the entire sentence containing that name. 3) Facts can include: attack tactics, techniques and procedures (TTPs), targets, motivations, industries, origin, date, etc. 4) Do not reference the original source, URL, or include any actor names other than the anonymized 'they'. 5) Output only the bullet list, no extra commentary.
\end{tcolorbox}

\subsection{Risk Mitigation Strategy Instruction}\label{appendix:systeminstructrms}

\begin{tcolorbox}[title = RMS System Instruction]
    You are an expert cyber threat analyst. Create a brief 2-3 sentence of specific attack scenario based solely on the attack pattern description below: Ensure the scenario reflects only this pattern and does not add extra techniques. Do not discuss ramifications or provide commentary. Return only the scenario text with no preface or postscript. You can learn from the three examples provided below:   
    
    \{EXAMPLES\}
\end{tcolorbox}

\begin{tcolorbox}[title = RMS Example]
\textbf{Example 1:}

\textit{attack-pattern-description:} Adversaries may abuse Windows Management Instrumentation (WMI) to execute malicious commands and payloads.

\textit{detail-attack-scenario:} An adversary remotely connected to a compromised workstation and used Windows Management Instrumentation (WMI) to execute a command that launched a malicious script. The script was delivered and run entirely through WMI without requiring the attacker to interact with the system's graphical interface.

\textbf{Example 2:}

\textit{attack-pattern-description:} Adversaries may attach filters to a network socket to monitor then activate backdoors used for persistence or command and control.

\textit{detail-attack-scenario:} An attacker gained access to a compromised server and attached a packet filter to a network socket to quietly observe inbound traffic. After monitoring activity for several hours, the adversary triggered a backdoor listener on the same socket to establish remote control of the system.

\textbf{Example 3:}

\textit{attack-pattern-description:} Adversaries may use utilities to compress and/or encrypt collected data prior to exfiltration.

\textit{detail-attack-scenario:} An attacker gathered sensitive project files from a compromised server and used a command-line utility to compress them into a single archive. To further secure the stolen data, the adversary applied password-based encryption to the archive before preparing it for transfer.
\end{tcolorbox}

\subsection{Evaluation Prompts}\label{appendix:evalprompts}

\begin{tcolorbox}[title=CKT Evaluation Prompt]
You are given a multiple-choice question (MCQ) from a Cyber Threat Intelligence (CTI) benchmark. Choose the single best option among the five provided (A–E). Briefly justify your choice. The final line of your response must begin with "Answer:" and contain only a single uppercase letter (A, B, C, D, or E) in the following exact format, without any extra commentary or text after it.

Final line example format:     

Answer: C

Question: \{\}
\end{tcolorbox}

\begin{tcolorbox}[title=RCM Evaluation Prompt]
Analyze the following CVE description and identify the most appropriate Common Weakness Enumeration (CWE) that represents the root cause of the vulnerability. Briefly justify your choice.\ The final line of your response must begin with "Answer:" and contain only the CWE ID in the following exact format, without any extra commentary or text after it.

Final line example format: 

Answer: CWE-287

CWE description: \{\}
\end{tcolorbox}

\begin{tcolorbox}[title=VSP Evaluation Prompt]
Analyze the following CVE description and calculate the CVSS v3.1 Base Score. Determine the values for each base metric: AV, AC, PR, UI, S, C, I, and A. For each metric, state the chosen value and briefly justify your choice. Valid options for each metric are as follows: Attack Vector (AV): Network (N), Adjacent (A), Local (L), Physical (P) Attack Complexity (AC): Low (L), High (H) Privileges Required (PR): None (N), Low (L), High (H) User Interaction (UI): None (N), Required (R) Scope (S): Unchanged (U), Changed (C) Confidentiality (C): None (N), Low (L), High (H)Integrity (I): None (N), Low (L), High (H)Availability (A): None (N), Low (L), High (H) After summarizing all metrics, construct the complete CVSS v3.1 vector string. The final line of your response must begin with "Answer:" and contain only the CVSS v3.1 Vector String in the following exact format, without any extra commentary or text after it.

Final line example format: 

Answer: CVSS:3.1/AV:N/AC:L/PR:N/UI:N/S:U/C:H/I:H
CVE Description: \{\}
\end{tcolorbox}

\begin{tcolorbox}[title=ATE Evaluation Prompt]
Analyze the following attack scenario in a Windows environment. Identify the single MITRE ATT\&CK technique (ID format: T\#\#\#\#) that best matches the behavior described. Briefly justify your choice. Only ATT\&CK techniques—do not output sub-techniques (T\#\#\#\#.\#\#\#), mitigations (M10xx), groups, software, or multiple IDs. The final line of your response must begin with "Answer:" and contain only the technique ID in the following exact format, without any extra commentary or text after it.

Final line example format: 

Answer: T1059

Attack scenario: \{\}
\end{tcolorbox}

\begin{tcolorbox}[title=TAA Evaluation Prompt]
Analyze the following anonymized incident summary, which includes attacker behaviors, tools, infrastructure, file/artifact usage, operational tradecraft, targeting, and other contextual details. Any direct mentions of the threat actor group have been anonymized with the pronoun “they”. Your task is to attribute the description to a known threat actor based on the techniques, tactics, procedures (TTPs), and any other relevant information described. Please provide the name of the threat actor you believe is responsible and briefly explain your reasoning.\ The final line of your response must begin with "Answer:" and contain only the threat actor in the following exact format, without any extra commentary or text after it.

Final line example format:     

Answer: Lazarus 

Threat report: \{\}
\end{tcolorbox}

\begin{tcolorbox}[title=RMS Evaluation Prompt]
Analyze the following attack scenario in a Windows environment. Recommend the most appropriate MITRE ATT\&CK mitigation ID(s) (format: M10xx) that address the scenario. Return exactly 4 mitigation IDs as a comma-separated list. Briefly justify your choice. Only ATT\&CK mitigations (M10xx)—do not output techniques (T\#\#\#\#), groups, or generic advice. The final line of your response must begin with "Answer:" and contain only the mitigation IDs in the following exact format, without any extra commentary or text after it.

Final line example format: 

Answer: M1037, M1041

Attack scenario: \{\}
\end{tcolorbox}

\end{document}